\documentclass[%
reprint,
amsmath,amssymb,
aps,
]{revtex4-2}

\usepackage{graphicx}
\usepackage{dcolumn}
\usepackage{bm}

\begin{document}
	
	\preprint{APS/123-QED}
	
	\title{Inconsistencies between experimental and theoretical band structure of TiSe$_2$}

	\author{Turgut Yilmaz}
	\affiliation{National Synchrotron Light Source II, Brookhaven National Lab, Upton, New York 11973, USA}
     \email{trgt2112@gmail.com}
	
	\author{Anil Rajapitamahuni}
	\affiliation{National Synchrotron Light Source II, Brookhaven National Lab, Upton, New York 11973, USA}
	
	\author{Elio Vescovo}
	\affiliation{National Synchrotron Light Source II, Brookhaven National Lab, Upton, New York 11973, USA}

	\date{\today}
	
	\begin{abstract}
		
		Renew interest in the charge density wave phase of TiSe$_2$ stems from the realization of its unique driving mechanism, the so called excitonic insulator phase. Existing claims are motivated by model calculations of the band structure. In this study, angle resolved photoemsision spectroscopy and density functional theory for TiSe$_2$ are directly compared. The substantial discrepancies found between the two descriptions cast serious doubts on the exitonic insulator scenario as the correct physical mechanism underlying the periodic lattice distortion at low temperature. In particular, the formation of a valence-conduction hybridization gap in the bulk band structure is not present in the experimental data. Therefore, the origin of the structural transition in TiSe$_2$ cannot be fully explained within the existing theoretical models.

	\end{abstract}

	\maketitle
	

Charge density wave (CDW) is an intriguing second order phase transition, usually occurring in low dimensional systems \cite{gruner1988dynamics, gor2012charge}. Below a critical temperature, the lattice experience a periodic distortion accompained by the opening of gaps at specific points of the Fermi surface (nesting sites), lowering the system energy \cite{gruner1988dynamics, gor2012charge}. Among the many materials, CDW is most frequently studied in transition metal dichalcogenides (TMDCs). TiSe$_2$ is a case in point, having gained particular attention due to emergent superconductivity with Cu doping or under pressure \cite{kusmartseva2009pressure,morosan2006superconductivity}. It undergoes a structural transition at 200 K with a formation of a 2a $\times$ 2a $\times$ 2c periodic lattice distortion \cite{stoffel1985experimental,di1976electronic}. However, unlike in other TMDC materials, the phase transition in TiSe$_2$ cannot be explained within the ordinary nesting scenario since the Fermi surface is claimed to have no parallel portions \cite{rossnagel2011origin,pillo2000photoemission}.

Although identifying the mechanism responsible to stabilize the distorted phase has been the subject of many studies, representing decades of cumulative efforts, the topic still remains elusive and controversial. So far, three main scenarios have been proposed. First is the excitonic insulator model as a purely electronic interaction \cite{pillo2000photoemission, cercellier2007evidence, monney2010probing}. In this model, the CDW is driven by the poorly screened Coulomb interactions between electrons and holes. A second model attributes to the electron-phonon coupling the origin of the CDW instability, which is eventually resolved via Peirls or Jahn-Teller type of mechanisms\cite{hughes1977structural}. Finally, in the third model, the combined effect of electron-hole and electron-phonon coupling is responsible for the transition \cite{van2010exciton}. 

In this context, the excitonic insulator phase has received the greatest attention \cite{monney2010temperature,kogar2017signatures,hellmann2012time,monney2012electron,cercellier2007evidence}. The identification of an emergent bulk band gap with a characteristic band dispersion in the distorted phase is a key element of this model. Therefore, angle resolved photoemission spectroscopy (ARPES) is the experimental tool of choice to verify this model. The prediction is based on the large overlap ($>$ 0.55 eV) between valence and conduction bands at the $\Gamma$ and $L$ points, respectively, found in the calculated band structure for the normal phase of TiSe$_2$.\cite{cercellier2007evidence,huang2021aspects,cazzaniga2012ab,kidd2002electron,monney2009spontaneous,bianco2015electronic,watson2019orbital}. Upon a periodic lattice distortion, these two high symmetry points become equivalent with consequent folding of the bands. The resulting band degeneracy close to the Fermi level would then be resolved by opening an hybridization gap, thereby leading to the realization of the excitonic insulator material.

   \begin{figure*}[t]
	\centering
	\includegraphics[width=16cm,height=8.213cm]{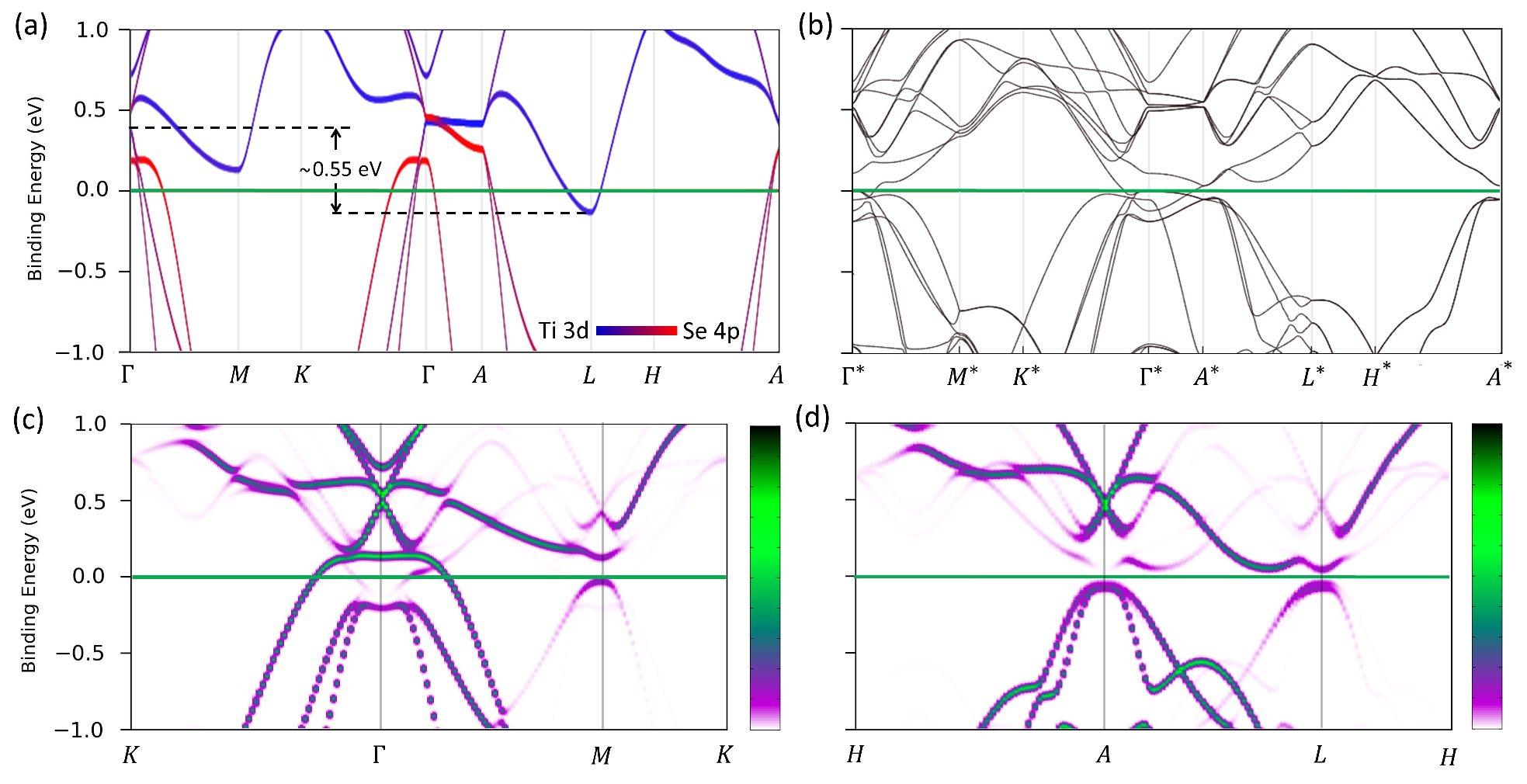}
	\caption{
		(a) Atomic orbitals projected band structure of TiSe$_2$ in the normal phase. 
		Red and blue represents the contributions from Se 4$p$ and Ti 3$d$ orbitals, respectively. Dashed lines mark the indirect bulk band gap of $\sim$ 0.55 eV. (b) Folded band structure for the $2a \times 2a \times 2c$ phase along the $\Gamma^* - M^* - K^* - \Gamma^* - A^* - L^* - H^* - A^*$ high symmetry directions of the supercell Brillouin zone. (c-d) Unfolded band structure along the $K - \Gamma - M - K$ and $H - A - L - H$ directions, respectively. Horizontal green lines represent the Fermi level obtained from DFT. Color contrast in (c-d) represents the spectral weight.
	}
\end{figure*}

\begin{figure*}[t]
	\centering
	\includegraphics[width=16cm,height=8.856cm]{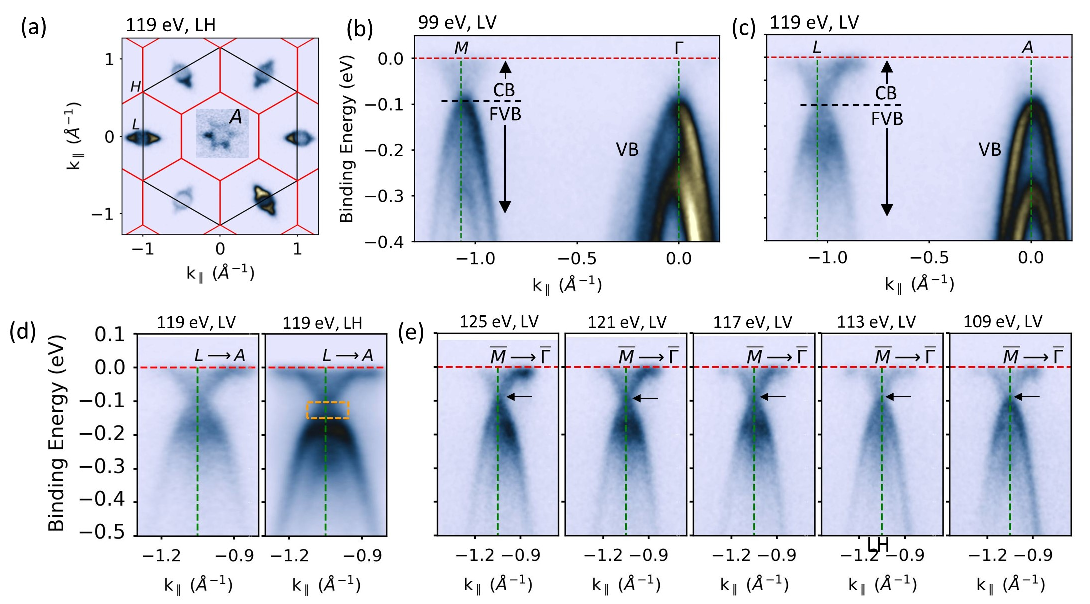}
	\caption{
		(a) ARPES Fermi surface at k$_z$ = $A$. The zone center is reproduced with higher color saturation to improve visibility of the band. (b-c) Electronic structures along the $M$ - $\Gamma$ and $L$ - $A$ directions respectively. FVB stands for folded valence bands. (d) Surface electronic structures at $L$-point taken with LV and LH polarized light. Dashed orange square marks the low spectral intensity region relative to the rest of the band structure (e-f) Photon energy dependent surface electronic structure at $\overline{M}$. Arrows in (e) point to the band overlap regions.
	}
\end{figure*}

 \begin{figure}[t]
 	\centering
 	\includegraphics[width=8.5cm,height=8.297cm]{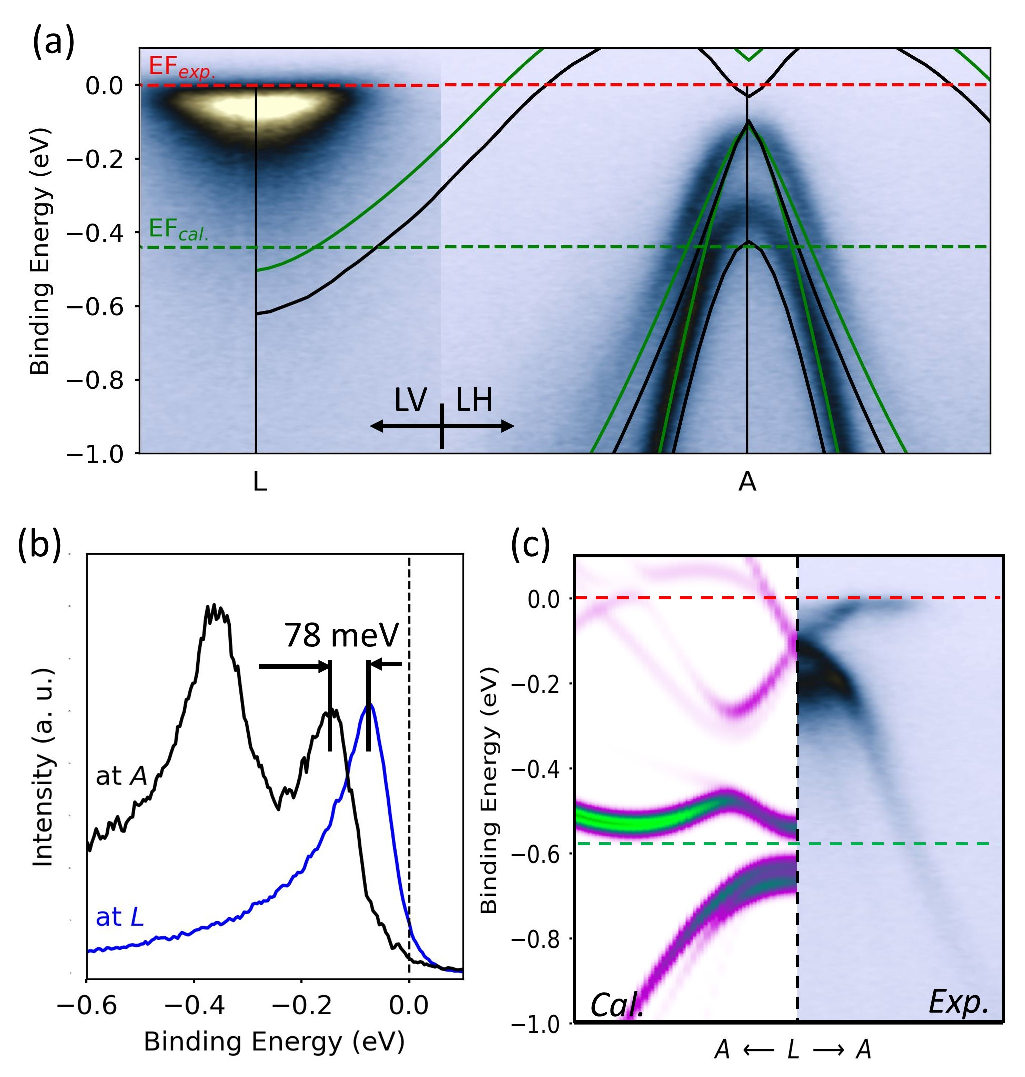}
 	\caption{
 		(a) Comparison of the experimental and computed band structure for the normal phase of TiSe$_2$, with (black) and without (green) spin orbit coupling. The calculations are down shifted about 600 meV to align with the measured valence band. Band structure around $L$-point is recorded with LV polarized light while the one around $A$ point is recorded with LH polarized light to enhance the spectral intensity. 119 eV photons are used for both sections. (b) EDCs along the $A$ and $L$ points to measure the indirect bulk band gap. (c) Side by side comparison of the computed and measured electronic structure in the distorted phase.}
 \end{figure}

\begin{figure}[t]
	\centering
	\includegraphics[width=8.5cm,height=8.34cm]{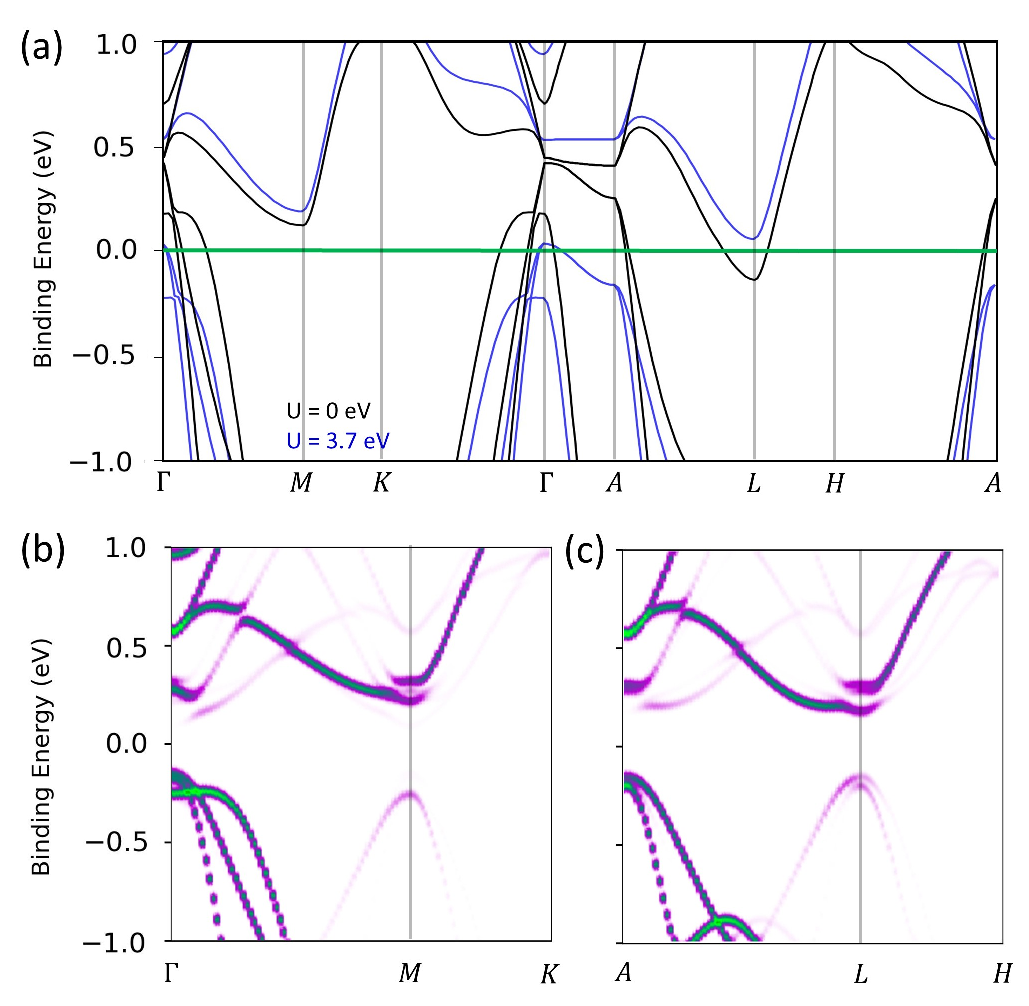}
	\caption{
             (a) Computed electronic structure with U = 3.7 eV (blue) and without U (black line). (b-c) Unfolded band structures along the $\Gamma -M - K$ and $A -L - H$ in the distorted phase by using U=3.7 eV, respectively. Horizontal green lines represent the Fermi level.}
\end{figure}

However, the predicted bands overlap is not born out by the experimental band structure of TiSe$_2$, which shows an indirect bulk band gap between valence band maximum and conduction band minimum in the normal phase \cite{chen2015charge,chen2016hidden,wegner2020evidence,watson2019orbital}. This would prevent band hybridization in the distorted phase which is the essential mechanism for the exciton insulator phase to emerge. But the gap is smaller than $\sim$ 50 meV, and therefore it is assumed that the overall predictions will not be affected \cite{rossnagel2011origin}. In principle, one could speculate on a temperature induced band shift to close the gap. However conduction and valence band are found experimentally to shift rigidly in the same direction as a function of temperature \cite{monney2010probing}. Therefore, conduction and valence band overlapping is extremely unlikely to occur and with it the band hybridization in the distorted phase.

Furthermore, our recent ARPES study shows that the experimental spectra are compatible with a gapless surface electronic structure of TiSe$_2$ \cite{yilmaz2023gapless}. In the same work, the earlier ARPES results are largely reproduced and new results are added, showing that the observation of a bulk band gap is based on undefined features in the ARPES spectra and is therefore dubious. Therefore, inconsistencies between the experimental and theoretical band structure remain elusive and lead to inconclusive interpretations of the experimental data. These issues are critical to understand the nature of the structural transition in TiSe$_2$.

Here, we present a combined photoemssion and density functional theory (DFT) study of the band structure of TiSe$_2$. The predicted and measured dispersion of the bands differs significantly. In particular, it is impossible to fit simultaneously the positions of valence and conduction bands. The expected band hybridization between conduction and valance band is absent in the experimental data. Furthermore, the surface electronic structure is found gapless in the distorted phase, directly ruling out the exitonic insulator scenario. Surprisingly, only the Fermi surface nesting remains as possible mechanism for the CDW transition.

ARPES experiments were performed at the 21ID-I ESM beamline at the National Synchrotron Light Source II (NSLS-II), using a DA30 Scienta electron spectrometer with an energy resolution better than 15 meV.  The analyzer slit was along to the $\overline{M}$ - $\overline{\Gamma}$ - $\overline{M}$ direction of the hexagonal Brillouin zone. The electric field with linear vertical (LV) polarized light was parallel to the sample surface and analyzer slit while laying in the incident plane for linear horizontal (LH) polarized light. Band structure calculations were performed with the Quantum Espresso (QE) package, based on density functional theory \cite{giannozzi2009quantum,giannozzi2017advanced}. Crystal structure for the distorted phase was adopted from Ref. 6. Samples were obtained from 2dsemicondcutors.

The orbital projected band structure for the normal phase is given in Figure 1(a). Spectral contribution to the valence band around the $\Gamma$($A$)-point is dominated by the Se-4p and the conduction band is contributed mostly by V-3d atomic orbitals. The Fermi level crosses the bottom of the conduction band at $L$ point while the tip of the valence band sits $\sim$ 0.46 eV above the predicted Fermi level. This leads to have a $\sim$ 0.55 eV overlapping between two bands as marked with dashed horizontal lines in Figure 1(a).

The folded electronic structure in the 2 $\times$ 2 $\times$ 2 distorted phase presents a strong overlap between the valence and conduction bands inducing a $p$ - $d$ band hybridization at the Fermi level (Figure 1(b)). In detail, an energy gap opens at the $L$-point while the band structure remains gapless at the $M$-point due to the orbital selective k$_z$ dispersion of the valence band \cite{watson2019orbital}. Furthermore, as shown in the unfolded band structure, the valence band tip flattens (Figure 1(c-d)). This modification in the electronic structure is widely regarded as the characteristic signature of the exitonic insulator state, manifesting through the formation of the back-folded valence and conduction bands \cite{monney2010temperature,kogar2017signatures,hellmann2012time,monney2012electron,cercellier2007evidence}.

We next discuss the experimental band structure in Figure 2, and 3. A well defined Fermi surface is observed at the $A$-plane. It consists of ellipsoidal pockets located at $L$-point and a star shaped dispersion in the zone center (Figure 2(a)). The periodic lattice distortion is confirmed through the observation of folded valence bands at the $M$ and $L$-points in the ARPES spectra (Figure 2(b-c)). In contrast to the expected bulk gap resulting from the $p$ - $d$ hybridization, clear gapless surface electronic structure is observed with a overlap at $\sim$ 100 meV below the Fermi level (Figure 2(d)). With linear horizontal line the situation is less clear, due to the higher intensity of a second, lower lying, valence band. The resulting intensity decrease just above this band can give the appearance of the opening of a gap between valence and conduction in the region marked by the yellow box. Furthermore, the band overlap displays a 2D nature as shown in the photon energy dependent data (Figure 2(e)). This behavior renders unlikely a k$_z$ dependent bulk band gap scenario as proposed in previous studies \cite{watson2019orbital,chen2016hidden}.

Experimental and computed band structures are compared in Figure 3 to disclose inconsistencies and find the origin of the gapless surface electronic structure. The energy scales are aligned at the top of the valence band at $A$-point (Figure 3(a)). Note that in this case there is a large mismatch between theory and experiment at the $L$-point. At variance with the large overlap predicted for the $p$-$d$ states, the experimental band structure in the normal state exhibits a 78 meV indirect gap between conduction and valence bands (Figure 3(a-b)) as consistent with previous reports \cite{watson2019orbital,chen2016hidden,chen2015charge}. Thereby, the band folding in the distorted phase is not expected to induce an hybridization gap.
A similar type of intractable inconsistency is found in the distorted phase. Persisting with the same energy scale alignment, one achieves a nice matching of bands close to the Fermi level at the expense of a complete failure at higher binding energies.

As shown above, calculated and experimental band structures are essentially consistent in the literature but differ considerably between themselves. In this situation to compare them, either the binding energy of the conduction band bottom or the tip of valence band needs to be taken as a reference. A criterion must be determined in order to decide which one is more relevant to the experimental data. In previous studies where the conduction band bottom is aligned, a prominent inconsistency for the rest of the band structure is evident \cite{bianco2015electronic,lian2020ultrafast}. Specifically, the major issue is that the valence band tip falls far above the Fermi level. Then, the indirect bulk band gap in the normal phase becomes inaccessible in ARPES while it is experimentally claimed to be persistent across the structural transition \cite{watson2019orbital,chen2016hidden,chen2015charge}. However in all experimental works, it is well established that the valence band is located below the Fermi level. Therefore, it seems more safe to take the binding energy of the valence band as a reference point to match the computed band structures.

The essential mismatch between measured and calculated band structure is known in the literature. But it is usually ignored when the experimental data are analyzed \cite{watson2019orbital,rossnagel2011origin,bianco2015electronic,hellgren2021electronic}. A few studies address the issue by incorporating an Hubbard U term and a quasi-self-consistent G0W0 approach \cite{bianco2015electronic,hellgren2021electronic}. However, the beneficial effects of these $ad-hoc$ methods in reproducing some aspect of the band structure are voided by the new problems that they bring in. We reproduce a similar band structure calculation using an Hubbard U term and obtain a narrow indirect bulk band gap of 25 meV in the normal phase (Figure 4). In this case, significantly larger bulk band gap of 0.4 eV is formed in the distorted phase, being much larger then any reported one in the literature. The inconsistency in this case is due to the underestimated band structure in the zone center. Adding a U-term to modify the conduction band makes TiSe$_2$ in the normal phase a narrow gap band insulator. Therefore, there is no Fermi surface to consider for nesting or band hybridization scenario. Hence, it is likely that not only the binding energy of the conduction band mismatch but also its dispersion is inconsistent with the experimental data. This probably the reason of having different velocities given in Figure 3(c) while gapless bulk band structure can be obtained in both experimental and theoretical data. Furthermore, Heyd-Scuseria-Ernzerhof (HSE06) functional is also adopted to reproduce experimental band structure of TiSe$_2$ in a more accurate way \cite{chen2016hidden,hellgren2017critical}. But this approach also introduces a large bulk band gap with the conduction band located far above the Fermi level.

In summary, we present a comparative study between experimental and theoretical band structure of TiSe$_2$ and show the inconsistencies between two. Particularly, the predicted band overlap between conduction and valence band is absent in the experimental data. Therefore, current findings rules out the possibility of an excitonic insulator phase as the mechanism behind the structural transition. The characteristic band dispersions of such phase, a bulk band gap with  back-folded conduction and valance band, are absent in the surface electronic structure. Contrary to the common belief, if the structural phase transition of TiSe$_2$ is treated as a CDW transition, the only possible driving mechanism remains a Fermi surface nesting. This conclusion can pave the way for future studies.

This research used resources ESM (21ID-I) beamline of the National Synchrotron Light Source II, a U.S. Department of Energy (DOE) Office of Science User Facility operated for the DOE Office of Science by Brookhaven National Laboratory under Contract No. DE-SC0012704. We have no conflict of interest, financial or other to declare.

\end{document}